\documentclass[english,prl,twocolumn, showpacs]{revtex4}
\usepackage[T1]{fontenc}
\usepackage[latin9]{inputenc}
\usepackage{amstext}
\usepackage{graphicx}

\makeatletter
\@ifundefined{textcolor}{}
{%
 \definecolor{BLACK}{gray}{0}
 \definecolor{WHITE}{gray}{1}
 \definecolor{RED}{rgb}{1,0,0}
 \definecolor{GREEN}{rgb}{0,1,0}
 \definecolor{BLUE}{rgb}{0,0,1}
 \definecolor{CYAN}{cmyk}{1,0,0,0}
 \definecolor{MAGENTA}{cmyk}{0,1,0,0}
 \definecolor{YELLOW}{cmyk}{0,0,1,0}
}

\makeatother

\usepackage{babel}
\begin{document}

\title{Quantification of Symmetry }

\author{Y. N. Fang$^{1,3,4}$, G. H. Dong$^{1,4}$, D. L. Zhou$^{2}$ and
C. P. Sun$^{1,4}$}

\email{cpsun@csrc.ac.cn}

\address{$^{1}$Beijing Computational Science Research Center, Beijing 100084,
China\\
$^{2}$Beijing National Laboratory for Condensed Matter Physics and
Institute of Physics, Chinese Academy of Sciences, Beijing 100190,
China\\
$^{3}$State Key Laboratory of Theoretical Physics, Institute of Theoretical
Physics, Chinese Academy of Sciences, and University of the Chinese
Academy of Sciences, Beijing 100190, China\\
$^{4}$Synergetic Innovation Center of Quantum Information and Quantum
Physics, University of Science and Technology of China, Hefei, Anhui
230026, China}
\begin{abstract}
Symmetry is conventionally described in a contrariety manner that
the system is either completely symmetric or completely asymmetric.
Using group theoretical approach to overcome this dichotomous problem,
we introduce the degree of symmetry (DoS) as a non-negative continuous
number ranging from zero to unity. DoS is defined through an average
of the fidelity deviations of Hamiltonian or quantum state over its
transformation group $G$, and thus is computable by making use of
the completeness relations of the irreducible representations of $G$.
The monotonicity of DoS can effectively probe the extended group for
accidental degeneracy while its multi-valued natures characterize
some (spontaneous) symmetry breaking. 
\end{abstract}


\maketitle

\paragraph{Introduction.\textemdash{}}

Symmetry is a theme of modern physics, which plays a crucial role
in the understanding of fundamental interactions of the microscopic
world \cite{2014 Sundermeyer BOOK} as well as the emergence of macroscopic
orders \cite{1984 Anderson BOOK}. It has become evident that both
the elementary particle structure and the emergent phenomena, e.g.,
superconductivity and Bose-Einstein condensation, are originated from
symmetry and its spontaneous breaking \cite{1956 Penrose PR,1961 Nambu PR,1991 Leggett}.
Its applications range from particle physics \cite{1964 Englert PRL,1964 Higgs PRL,2014 Kibble PTRSA,2004 Witten Nature}
to condensed matter physics \cite{2015 Navon Science,2014 Liu Nat Commun},
and even to biological systems \cite{2015 Duboc Ann Rev,2013 Saito RMP}.

Conventionally, symmetry is dealt in a dichotomous fashion that a
physical system either possesses or not possesses a symmetry. In the
group theoretical approach, the symmetry of a quantum system is usually
considered by checking that if the system is invariant or not under
some transformations, which sometimes form a symmetry group $G$.
The symmetry breaking of the system can be described as a reduction
of the symmetry group to its subgroup. Although this conventional
approach has succeed in classifying the spectrum structure and even
various phases of matters, it is not natural for us because there
is not a room for the intermediate circumstance, namely, a continuous
measure of symmetry has not been found. Actually, such intermediate
issues exist objectively and needs to be properly quantified. For
example, a charged particle moving in a central potential possesses
SO(3) symmetry. When a static magnetic field is applied, no matter
how weak it is, the SO(3) symmetry is said to be broken into SO(2).
However, SO(3) symmetry can still be approximately used to simplify
the equations describing the dynamics and the energy level structure
when the magnetic field is weak enough. Another example is the nuclear
system that possesses the isospin SU(2) symmetry and thus its energy
spectrum of strong interaction can approximately, but effectively,
be classified, although the electromagnetic force could break this
SU(2) symmetry. 

In this regard, it would be of much interest to present a quantitative
description of symmetry and its (spontaneous) breaking in this intermediate
circumstance, which could determine the extent of approximation for
using a given symmetry in practice. To this end, we, in this letter,
introduce a continuous measure of symmetry, i.e., the degree of symmetry
(DoS), by considering that symmetry is a relative concept: the particular
subset $G$ of all physically-allowed transformations needs to be
specified for assigning a symmetry to a physical system. More specifically,
for a given set $G$ of transformations on the Hamiltonian or the
quantum state $F=H$, or $\rho$, we first define a dual of DoS, the
degree $A(G,F)$ of asymmetry (DoAS), by averaging the fidelity deviations
(see definition below) over $G$. Generally, the DoAS ranges from
zero to unity, and thus the DoS $S(G,F)=1-A(G,F)$ also satisfies
$0\leq S(G,F)\leq1$. Evidently, $S(G,F)$ offers symmetry an intermediate
description to avoid the dichotomy in the conventional group theoretical
analysis. 

We will show that, if we chose $G$ as a group, the DoS , bounded
with $1/2\leq S(G,F)\leq1$, facilitates a general computable measure
of symmetry based on the irreducible representations of $G$. It is
potential in identifying various natures of symmetry that are important
to emergent phenomena, such as the spontaneous symmetry breaking (SSB).
For example, the thermodynamic SSB corresponds to multi-valued natures
of DoS at the low temperature, which is similar to the depiction of
the spontaneous magnetization \cite{1987 Huang BOOK}. It is also
shown that the multi-level crossing by a proliferation of energy levels
brings a peak to the DoS and the extended group can be given to account
for the hidden symmetry from accidental degeneracy.

\paragraph{Degree of symmetry.\textemdash{}}

We consider a quantum system with Hamiltonian $H$, and a set $G$
of $n_{G}$ transformations on its Hilbert space $\mathcal{H}$. When
$OHO^{-1}=H$ for $O\in G$, we say that $H$ (the quantum system)
is symmetric with respect to the transformation $O$. Actually, all
symmetric transformations form a group $G'(\subset G)$. It is obvious
that, the deviations of $OHO^{-1}$ from $H$ measure the extent of
the asymmetry of $H$ with respect to the transformation set. Thus,
we use their average over $G$ to define the degree of symmetry breaking
(asymmetry) DoAS
\begin{equation}
A(G,H)=\frac{1}{4|\tilde{H}|{}^{2}}\overline{|[R(g),H]|^{2}}
\end{equation}
where $|O|=\sqrt{\mathrm{Tr}\{O^{\dagger}O\}}$ indicates the Frobenius
norm \cite{2015 Watrous BOOK} while $\overline{f(g)}|_{G}\equiv\overline{f(g)}=n_{G}^{-1}\sum_{g\in G}f(g)$
is an average of a (group) function $f(g)$ defined on $G$, and later
the subscript $G$ will be occasionally omitted; If $G$ is a group,
then $R:\mbox{ }g\rightarrow R(g)\in End(\mathcal{H})$ is a \emph{d}-dimensional
representation of $g\in G$. Otherwise, $R(g)$ represents a unitary
transformation on $\mathcal{H}$. Here, $|[R(g),H]|^{2}=|R(g)^{\dagger}HR(g)-H|^{2}$
is the fidelity deviation of $H$ under the action of $g$, and $\tilde{H}=H-d^{-1}\mathrm{Tr}\{H\}$
is a re-biased Hamiltonian such that it is invariant under the zero-point
energy shifting $H\rightarrow H+\epsilon$ for $\epsilon$ being a
real number.

It is easy to prove that $0\leq A(G,H)\leq1$, thus the DoS defined
by $S(G,H)=1-A(G,H)$ or 
\begin{equation}
S(G,H)=\frac{1}{4|\tilde{H}|^{2}}\overline{|\{R(g),\tilde{H}\}|^{2}}\label{DoS}
\end{equation}
ranges from zero to unity and thus quantifies the extent of the symmetry
of $H$ with respect to $G$. The above definition of DoS is evidently
reasonable in physics since it possesses the following properties
(for the proofs see the supplemental material \cite{supplemental material}):
(1) Tighter bound when $G$ forms a transformation group $0\leq A(G,H)\leq1/2\leq S(G,H)\leq1$;
(2) Independence of DoS on the basis, i.e, $S(WGW^{\dagger},WHW^{\dagger})=S(G,H)$,
where $W$ is a unitary transformation and $WGW^{\dagger}=\{WR(g)W^{\dagger}|g\in G\}$;
(3) Scaling invariance, i.e., $S(G,\lambda H)=S(G,H)$; (4) Independence
of the choice of the zero-point energy, i.e., $S(G,H+\epsilon)=S(G,H)$;
(5) Hierarchy property $n_{G'}S(G_{s},H)\leq n_{G}S(G,H)$ for a subset
$G_{s}(\subset G\text{\ensuremath{)}}$ with $n_{G'}$ elements. 

When $G$ becomes a group, in the spaces $\mathcal{H}^{(l)}$ of its
$l$th irreducible representations with finite dimensions $d_{l}$,
the DoS $S=S(G,H)$ is re-expressed as
\begin{equation}
S=\frac{1}{2}+\sum_{l}\frac{1}{2d_{l}}(\sum_{\alpha}\frac{\langle l,\alpha|H|l,\alpha\rangle}{|\tilde{H}|}-\frac{\mathrm{Tr}\{H\}d_{l}}{d|\tilde{H}|})^{2},\label{DoS under irr-basis}
\end{equation}
where $|l,\alpha\rangle$ is a basis vector of $\mathcal{H}^{(l)}$
($\alpha=1,2,...,d_{l}$), and we have used the completeness relations
of irreducible representations \cite{supplemental material}. The
bisection point 1/2 from property (1) is also reflected in above equation,
since each term contributes non-negatively in the summation over $l$.
We point out that, by using Eq.(\ref{DoS under irr-basis}), DoS is
feasible to be computed based on the measurements with respect to
the basis $\{|l,\alpha\rangle\}$. Otherwise, for a continuous group,
a straightforward calculation of DoS from Eq.(\ref{DoS}) should need
to carry out the group integral with the Haar measure, e.g., the sum
over SO(3) becomes a Lie group integral \cite{supplemental material}.

\paragraph{Symmetry breaking.\textemdash{}}

Let $G$ be a symmetry group of the quantum system with Hamiltonian
$H$. A perturbation $H'=\lambda V$ breaks the symmetry into the
subgroup $G_{s}(\subset G)$, i.e., $[V,R(g)]\ne0$ for $g\in G-G_{s}$
and $[V,R(g')]=0$ for $g'\in G_{s}$. For the total Hamiltonian $H(\lambda)=H+\lambda V$,
we calculate the DoS under the symmetry breaking \cite{supplemental material}
\begin{equation}
S(G,H(\lambda))=1-\frac{A(G,V)\lambda^{2}}{\lambda^{2}+\xi\lambda+\eta},\label{DoS symmetry breaking}
\end{equation}
where $A(G,V)$ is the DoAS of the Hermitian operator $V$, the other
two coefficients are defined as $\xi=2\mathrm{Tr}\{\tilde{H}\tilde{V}\}|\tilde{V}|^{-2}$
and $\eta=|\tilde{H}|{}^{2}|\tilde{V}|{}^{-2}$. The above equation
exactly reflects the duality between symmetry and asymmetry: the maximal
symmetry breaking due to the perturbation corresponds to the minimal
symmetry of the considered system. When $|\lambda|$ is increased,
there exists a special point $\lambda_{\mathrm{A}}=-2\xi^{-1}\eta$
where the DoS reaches a local minimum $S_{\mathrm{min}}=1-A(G,V)\csc^{2}\varphi$;
here $\varphi$ is the angle between $\tilde{H}$ and $\tilde{V}$
\cite{supplemental material}.

\begin{figure}
\begin{centering}
\includegraphics[scale=0.24]{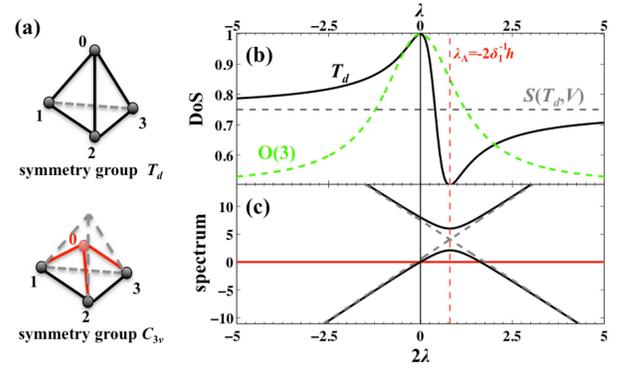}
\par\end{centering}

\caption{(color online). (a) Top: Schematic of a four-site lattice arranged
into the regular tetrahedron geometry, with Hamiltonian $H$ and symmetry
group $T_{d}$. Bottom: The $T_{d}$ symmetry is broken into $C_{3v}$
upon adding the perturbation $H'=\lambda V$, which changes the hopping
strength as well as the site energy relevant for the 0th site. (b)
Degree of symmetry (DoS) vs $\lambda$ for the four-site model (black
solid) and the angular momentum model (green dashed). The asymptotic
value $S(T_{d},V)$ (black dashed) and the local minimum $\lambda_{A}$
(red vertical line) for the four-site model are also shown. (c) Energy
spectrum of the four-site model vs $2\lambda$. Red line indicates
the two degenerate $E$ levels. Avoid level crossing of the two $A_{1}$
levels is shown by the grey dashed lines.}
\end{figure}
The following two examples are used to illustrate the above conception
on quantifying the extent of symmetry and its breaking. First, let
us consider a particle residing on a four-site lattice with the following
Hamiltonian
\begin{equation}
H=\mbox{\ensuremath{\sum}}_{i}\epsilon|i\rangle\langle i|+\mbox{\ensuremath{\sum}}_{ij}h|i\rangle\langle j|,\label{Four site Hamiltonian H}
\end{equation}
where $|i\rangle$ ($i=0,1,2,3$) is the single particle state with
site $i$ occupied. The site energy $\epsilon$ and the hopping strength
$h$ are site-independent for the regular tetrahedron geometry {[}see
Fig. 1a{]}, and thus $H$ is symmetric to all transformations from
the $T_{d}$ group, which contains (combined) rotations and mirror
reflections sending a regular tetrahedron into itself \cite{1964 Tinkham BOOK}.
In this example, we let the symmetry $T_{d}$ break into $C_{3v}$
through the following perturbation 
\begin{equation}
H'=\lambda[\delta_{0}|0\rangle\langle0|+\delta_{1}\mbox{\ensuremath{\sum}}_{i=1}^{3}(|i\rangle\langle0|+h.c.)],\label{Four site Hamiltonian V}
\end{equation}
where $\lambda\delta_{0}$ and $\lambda\delta_{1}$ are the deviations
of the energy and the coupling related to the 0th site. It is well
known that $C_{3v}$ has two one-dimensional irreducible representations
$A_{1}$ and $A_{2}$, as well as one two-dimensional irreducible
representation $E$ \cite{1964 Tinkham BOOK}, which correspond to
the three kinds of energy levels with one or two-fold degeneracies.

The above symmetry breaking from $T_{d}$ to $C_{3v}$ is quantified
by the DoS through Eq.(\ref{DoS symmetry breaking}) with $G=T_{d}$.
Straightforward calculation shows exact results $A(T_{d},V)=(2\gamma^{2}+16)^{-1}(\gamma^{2}+4)$,
$\xi=16(\gamma^{2}+8)^{-1}\delta_{1}^{-1}h$, and $\eta=\xi\delta_{1}^{-1}h$.
Here, $\gamma=\delta_{1}^{-1}\delta_{0}$ is the ratio between the
two parameters in $H'$. As shown in Fig. 1b, the DoS reaches unity
when $\lambda=0$, indicating the full $T_{d}$ symmetry that possessed
by the original Hamiltonian $H$. The symmetry breaking perturbation
$H'$ suppresses the DoS first quadratically in $\lambda$ and then,
as $|\lambda|$ further increased to approach the strong perturbing
limit ($|\lambda|\rightarrow\infty$), reaches a $\gamma$-dependent
asymptotically value $(2\gamma^{2}+16)^{-1}(\gamma^{2}+12)$. 

In this model, the special point $\lambda_{\mathrm{A}}=-2\delta_{1}^{-1}h$,
where the DoS reaches the local minimum, indicates an avoid level
crossing in the energy spectrum. To see this, we rewrite $H(\lambda)$
in terms of the standard basis of irreducible representations by using
the projection operator method \cite{1964 Tinkham BOOK,2011 Maze}.
The resulting four-dimensional Hilbert space contains two $A_{1}$-representations
and one $E$-representation of $C_{3v}$ \cite{1964 Tinkham BOOK}.
The two levels that transform according to the two $A_{1}$-representations
are coupled and the corresponding avoid level crossing point $\lambda_{*}$
is related to $\lambda_{\mathrm{A}}$ by
\begin{equation}
\lambda_{*}=\frac{6-\gamma}{12+\gamma^{2}}\lambda_{\mathrm{A}}.
\end{equation}
Especially, for $\delta_{0}\ll\delta_{1}$ the avoid level crossing
happens approximately at $\lambda_{\mathrm{A}}/2$ {[}see Fig. 1b,c{]}.

Another example demonstrates the DoS of the breaking of the continuous
symmetry. The system we considered is a particle with angular momentum
$j$, whose Hamiltonian reads
\begin{equation}
H=\epsilon J^{2},\mbox{ }H'=\lambda J_{z},\label{SO 3 model Hamiltonian}
\end{equation}
where $J_{i}$ ($i=x,y,z$) are components of the angular momentum
operator and $J^{2}=J_{x}^{2}+J_{y}^{2}+J_{z}^{2}$. In this model,
the O(3) symmetry of $H$ is broken by the perturbation, described
by $H'$, to O(2). With $G=\mathrm{O(3)}$, the DoS is calculated
as $1-[2\lambda^{2}+\epsilon^{2}j(j+2)]^{-1}\lambda^{2}$ \cite{supplemental material}.
Unlike the previous model, here the DoS does not show a local minimum
and decays monotonically as $\left|\lambda\right|$ increasing. Comparison
with the generic result Eq.(\ref{DoS symmetry breaking}) indicates
the underlying condition $\mathrm{Tr}\{\tilde{H}\tilde{V}\}=0$, which
is fulfilled by the Hamiltonian Eq.(\ref{SO 3 model Hamiltonian}).

\paragraph{Accidental degeneracy.\textemdash{}}

Accidental degeneracy of energy levels appears in a quantum system
when its parameters are changed to cause a level crossing. It is usually
not relevant to the geometric symmetry, but our DoS can reveal the
existence of the hidden symmetry. Actually, accidental degeneracy
also implies symmetry. The greater the degeneracy, the greater the
symmetry.

\begin{figure}
\begin{centering}
\includegraphics[scale=0.225]{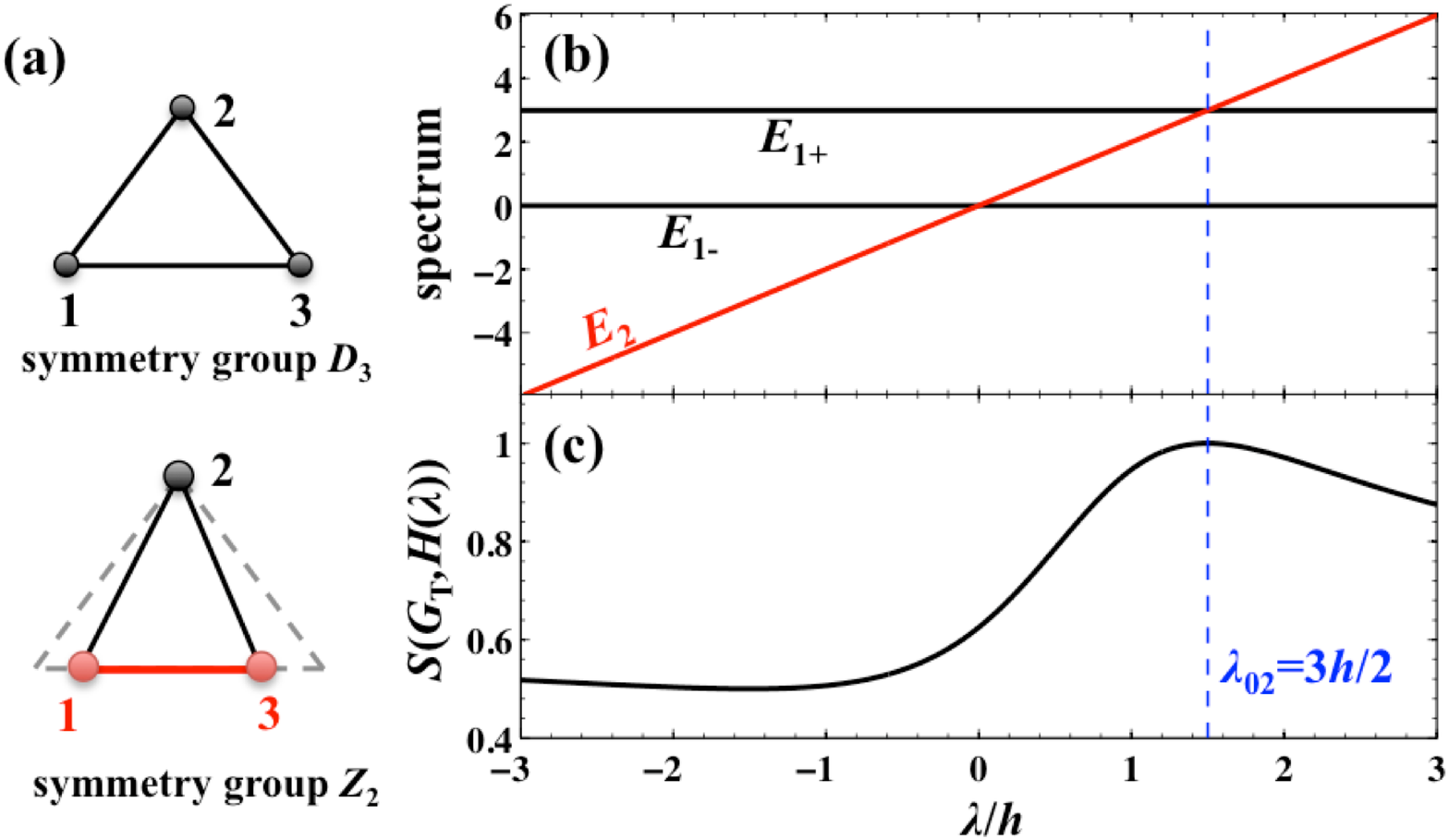}
\par\end{centering}

\caption{(color online). (a) Schematics of the three-site model, with symmetry
$D_{3}$ breaking into $Z_{2}$ by the perturbation Eq.(\ref{perturbation Hamiltonian three site model-1}).
(b) Energy spectrum vs $\lambda/h$ for the three-site model, showing
two accidental degeneracies at $\lambda_{01}$ and $\lambda_{02}$
(blue vertical line). (c) DoS vs $\lambda/h$ with respect to $G_{\mathrm{T}}$,
showing that the accidental degeneracy at $\lambda_{02}$ is identified
with the maximum of the DoS.}
\end{figure}

For the general Hamiltonian $H(\lambda)$ defined above, we introduce
the additional transformations: the $\mathrm{U}(2)$ operations on
the two $\lambda$-dependent energy levels of $H(\lambda)$ (or $\mathrm{U}(N)$
operations for the more general $N$ levels crossing), which will
become degenerate as $\lambda$ tuned to $\lambda_{0}$. Because $H(\lambda_{0})$
is proportional to the identity operator in the degenerate subspace
and, as a result, commuted with all $\mathrm{U}(2)$ operations, the
symmetry group $G$ of $H(\lambda_{0})$ is extended to a larger one
$G_{\mathrm{T}}=\langle G,\mathrm{U}(2)\rangle,$ which is generated
by elements in $G$ and $\mathrm{U}\text{(2)}$. It is expected that
the behavior of DoS could manifest the hidden symmetry that implied
by the enlarged group $G_{\mathrm{T}}$: the level crossing at $\lambda_{0}$
could result in a local dip in the DoAS, when the parameter $\lambda$
is tuned close to $\lambda_{0}$. To see this, we expand the Hamiltonian
linearly around $\lambda_{0}$, i.e., $H(\lambda)\approx H(\lambda_{0})+\partial_{\lambda}H(\lambda_{0})(\lambda-\lambda_{0})$.
Since $[R(g),H(\lambda_{0})]=0$ for $g\in G_{\mathrm{T}}$, the DoAS
is written as 
\begin{eqnarray}
A(G_{\mathrm{T}},H(\lambda)) & \propto & A(G_{\mathrm{T}},\partial_{\lambda}H(\lambda_{0}))(\lambda-\lambda_{0})^{2}.\label{DoAS argument-1}
\end{eqnarray}
Thus, by the duality, the accidental degeneracy indeed manifests itself
as a local maximum at $\lambda_{0}$ in DoS. 

To illustrate the above idea, we consider the following three-site
model whose Hamiltonian is of the same form as Eq.(\ref{Four site Hamiltonian H})
except that $i\in\{1,2,3\}$. And the perturbation term
\begin{equation}
H'=\lambda[|1\rangle\langle1|+|3\rangle\langle3|-(|1\rangle\langle3|+h.c.)]\label{perturbation Hamiltonian three site model-1}
\end{equation}
breaks the symmetry from $D_{3}$ to $Z_{2}=\{e,\sigma\}$. Here,
the transformation $\sigma$ interchanges the basis state $|1\rangle$
with $|3\rangle$. The energy spectrum of $H(\lambda)$ contains two
$\Gamma_{1}$ levels $E_{1\pm}=\epsilon+h/2\pm\lambda_{02}$ and one
$\Gamma_{2}$ level $E_{2}=\epsilon-h+2\lambda$, where $\Gamma_{i=1,2}$
are two irreducible representations of $Z_{2}$. The spectrum shows
two accidental degeneracies between the $\Gamma_{1}$ and the $\Gamma_{2}$
levels at $\lambda_{01}=0$ and $\lambda_{02}=3h/2$, respectively
{[}see Fig. 2b{]}. 

Indeed, at the accidental degeneracy, $H(\lambda_{02})$ becomes more
symmetric since there exists the additional symmetric transformations
of $\mathrm{U}(2)$: $R(\omega_{0};\hat{n},\omega)=\exp[i(\omega_{0}-\hat{n}\cdot\vec{s}\omega)]$
with pseudo spin-1/2 operators $\vec{s}$ defined by $s_{x}=(|\psi_{1+}\rangle\langle\psi_{2}|+|\psi_{2}\rangle\langle\psi_{1+}|)/2$
\emph{et al}.. Here, $|\psi_{m}\rangle$ is the eigenstate associated
with level $E_{m}$. The extended symmetry group $G_{\mathrm{T}}$
for $H(\lambda_{02})$ is still $\mathrm{U}(2)$ since $Z_{2}\subset\mathrm{U}(2)$
\cite{supplemental material}. Thus, the two-fold degenerate subspace
supports a two-dimensional irreducible representation of $G_{\mathrm{T}}$.
It is shown that the DoS $S(G_{\mathrm{T}},H(\lambda))$= $1-3[\lambda^{2}-\lambda_{02}\lambda+\lambda_{02}^{2}]^{-1}(\lambda-\lambda_{02})^{2}/8$
reaches the unity when $\lambda=\lambda_{02}$ {[}see Fig. 2c{]}.
Therefore, the DoS indeed signals the hidden symmetry. We notice that,
without the geometric symmetry, the above $\mathrm{U}(2)$ symmetry
defined in the subspace spanned by $|\psi_{1+}\rangle$ and $|\psi_{2}\rangle$
at the accidental degenerate point is similar to the dynamical symmetry
$\mathrm{SO}(4)$ of the non-relativistic hydrogen atom \cite{2011 Sakurai BOOK}.

\paragraph{DoS of quantum state and spontaneous symmetry breaking.\textemdash{}}

In emergent phenomena, the symmetry of the system ground state can
be different from that of the underlying Hamiltonian or Lagrangian.
This difference is roughly regarded as the spontaneous symmetry breaking
(SSB) \cite{1961 Nambu PR,1964 Higgs PRL,1984 Anderson BOOK}. For
a better depiction of those phenomena, we need to introduce the DoS
of quantum state (DoSS) $\rho$, which is analog to the DoS of Hamiltonian
by the Eq.(\ref{DoS})
\begin{equation}
S(G,\rho)=\frac{1}{4|\rho|^{2}}\overline{|\{R(g),\rho\}|^{2}}\label{DoS for quantum state}
\end{equation}
where $\rho$ is the density matrix of a quantum state. It possesses
the similar properties (1)-(5) except for $S(G,\rho)=S(G,\rho+\epsilon)$,
which we need not to require for physics.

\begin{figure}
\begin{centering}
\includegraphics[scale=0.225]{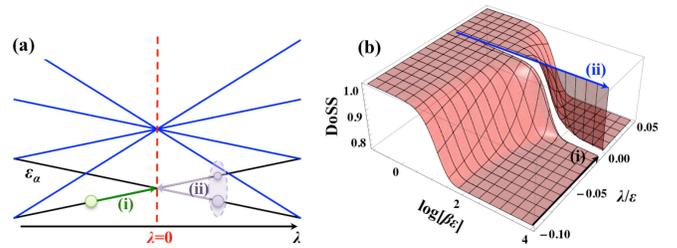}
\par\end{centering}

\caption{(color online). (a) Energy spectrum of a system with degenerate ground
states $\{|G_{\alpha}\rangle\}$ at $\lambda=0$. Upon whether $T\rightarrow0$
(i) before or (ii) after $\lambda\rightarrow0$, the thermal equilibrium
state Eq.(\ref{state in SSB}) approaches different final states.
(b) DoS of quantum state (DoSS) $S(\mbox{O(3)},\rho)$ vs $\lambda/\epsilon$
and $\beta\epsilon$ for the angular momentum model. The multi-valued
natures of the DoSS at $T=0$ and $\lambda=0$ are reflected as the
two non-commuting limiting processes indicated by (i) and (ii).}
\end{figure}
We now use DoSS to characterize the SSB in thermodynamics. We consider
the thermalization of a quantum system with degenerate ground states
$\{|G_{\alpha}\rangle|\alpha=1,2,...,d_{G}\}$, i.e., $H|G_{\alpha}\rangle=\varepsilon_{0}|G_{\alpha}\rangle$
\cite{2009 Liao arXiv}. At the zero temperature such system will
have a non-vanishing entropy $S=k_{\mathrm{B}}\ln d_{G}$, known as
the modified third law of thermodynamics \cite{1987 Huang BOOK}.
By introducing a perturbation $H'=\lambda V$ to break the symmetry
so that $|G_{\alpha=0}=G_{0}\rangle$ becomes the unique ground state,
the thermodynamic SSB is described as the following two non-commutative
limiting processes: (i) $T\rightarrow0$ and then $\lambda\rightarrow0$;
(ii) $\lambda\rightarrow0$ and then $T\rightarrow0$. In these two
non-commutative limiting processes, the following state
\begin{equation}
\rho=\frac{1}{Z}\mbox{\ensuremath{\sum}}_{\alpha\ne0}e^{-\varepsilon_{\alpha}/T}|G_{\alpha}\rangle\langle G_{\alpha}|+\frac{1}{Z}e^{-\varepsilon_{0}/T}|G_{0}\rangle\langle G_{0}|+...\label{state in SSB}
\end{equation}
will approach to $\rho_{\mathrm{f}1}=|G_{0}\rangle\langle G_{0}|$
and $\rho_{\mathrm{f2}}=d_{G}^{-1}\sum_{\alpha}|G_{\alpha}\rangle\langle G_{\alpha}|$
respectively {[}see Fig. 3a{]}.

To see the quantitative details of such thermodynamic SSB, we use
the DoSS defined by Eq.(\ref{DoS for quantum state}). Let $G$ be
a symmetry group of $H$, and $\{|G_{\alpha}\rangle\}$ span an irreducible
representation of $G$. Because $\rho_{\mathrm{f2}}$ is proportional
to the identity operator, thus $[R(g),\rho_{\mathrm{f2}}]=0$. This
implies that the limiting process (ii) results in a final state with
$S(G,\rho_{\mathrm{f2}})=1$. On the other hand, from Schur's theorem
\cite{1964 Tinkham BOOK}, in the limiting process (i) there always
exists some $g\in G$ such that $[R(g),\rho_{\mathrm{f1}}]\ne0$ and
consequently $S(G,\rho_{\mathrm{f1}})<1$. This in turn implies an
SSB since the final state $\rho_{\mathrm{f1}}$ does not retain the
full symmetry of the underlying microscopic Hamiltonian $H$.

The above relation between the DoSS and $\rho_{\mathrm{f}1,2}$ suggests
the multi-valued natures of DoSS at ($T=0$, $\lambda=0$) upon different
limiting processes as an SSB witness 
\begin{equation}
\lim_{\beta\rightarrow\infty}\lim_{\lambda\rightarrow0}S(G,\rho)=1,\mbox{ }\lim_{\lambda\rightarrow0}\lim_{\beta\rightarrow\infty}S(G,\rho)<1,\label{SSB witness}
\end{equation}
where $\beta=1/T$ is the inverse temperature.

To illustrate the SSB with an example, let us consider the angular
momentum model Eq.(\ref{SO 3 model Hamiltonian}) again, which shows
a spontaneous breaking of O(3) symmetry. In the subspace with $j=1/2$,
the ground state is two-fold degenerate without $H'$, i.e., $|1/2,\pm1/2\rangle$.
For a generic thermal state $\rho=Z^{-1}\exp[-\beta H(\lambda)]$,
DoSS is shown to be $S(\mbox{O(3)},\rho)=(3+\cosh^{-1}\beta\lambda)/4$
\cite{supplemental material}, whose multi-valued natures at ($T=0$,
$\lambda=0$) is shown in Fig. 3b. Specifically, when $\beta\epsilon\gg1$
while $\lambda\ne0$ the DoSS is nearly at a constant value 3/4. Then,
by tuning the coupling $\lambda$ to zero, the DoSS remains fixed
at the same constant value (see (i) in Fig. 3b). In contrast, if one
first fix the coupling $\lambda=0$ at the high temperature, the DoSS
as a function of $\beta$ from zero to infinity will follow the blue
arrowed line (corresponding to the possess (ii)) in Fig. 3b. In the
latter case, the DoSS at large $\beta$ is unity. In analog to Eq.(\ref{SSB witness}),
in this example it is shown that
\begin{equation}
\lim_{\beta\rightarrow\infty}\lim_{\lambda\rightarrow0}S(G,\rho)=1,\mbox{ }\lim_{\lambda\rightarrow0}\lim_{\beta\rightarrow\infty}S(G,\rho)=\frac{3}{4}.
\end{equation}
Here, $G=\mathrm{O}(3)$. On the other hand, the difference in DoSS
reflected by above equations is also understood through inspecting
on the final state, which is $\rho_{\mathrm{f1}}=\left|1/2,-1/2\left\rangle \right\langle 1/2,-1/2\right|$
or $\rho_{\mathrm{f2}}=2^{-1}\sum_{m}|1/2,m\rangle\langle1/2,m|$
upon limiting processes (i/ii). Clearly, $\rho_{\mathrm{f1}}$ is
not invariant under the $\pi$-rotation that represented by the $\sigma_{x}$
operation in the $j=1/2$ subspace, thus results in a DoSS smaller
than unity.

\paragraph{Conclusion.\textemdash{}}

In this letter, we introduce a continuous measure, the degree of symmetry
(DoS), for the symmetry of quantum system, which largely extrapolates
the dichotomous approach of symmetry based on group representation
theory. It is shown that the DoS possesses some good properties, such
as basis-independent, invariant under the zero-point energy shifting
as well as the scaling transformation. Since it can be expressed as
an average of physical operators under the basis of irreducible representations
for transformation groups, this measure is thus computable and detectable
based on some quantum measurements. 

In contrast to the previous explorations based on the abstract concepts
of fuzz set \cite{2010 Garrido Symmetry-1} and transform information
\cite{1997 Vstovsky Found Phys-1}, our introduced DoS can feasibly
open many applications in physics. As illustrated in this letter,
the DoS is capable of identifying symmetry relevant phenomena and
effects, such as the accidental level crossings and the spontaneous
symmetry breaking. This, therefore, implies that the DoS could be
a useful measure in related future studies, e.g., in characterizing
systems near quantum criticality since it is closely related to the
multi-level crossings \cite{2006 Quan PRL}. 
\begin{acknowledgments}
We thank X. F. Liu, P. Zhang, X. G. Wang, S. X. Yu, and L. P. Yang
for helpful discussions. This work was supported by the National Natural
Science Foundation of China (Grant Nos. 11421063, 11534002, and 11475254)
and the National 973 program (Grant Nos. 2014CB921403, 2012CB922104,
and 2014CB921202). \end{acknowledgments}

\end{document}